\documentstyle[emulateapj,graphicx]{article}
\def\simlt{\lower.5ex\hbox{$\; \buildrel < \over \sim \;$}}
\def\simgt{\lower.5ex\hbox{$\; \buildrel > \over \sim \;$}}
\def\gsim{\;\rlap{\lower 2.5pt
 \hbox{$\sim$}}\raise 1.5pt\hbox{$>$}\;}
\def\lsim{\;\rlap{\lower 2.5pt
   \hbox{$\sim$}}\raise 1.5pt\hbox{$<$}\;}

\def\spose#1{\hbox to 0pt{#1\hss}}
\def\lta{\mathrel{\spose{\lower 3pt\hbox{$\mathchar''218$}}
     \raise 2.0pt\hbox{$\mathchar''13C$}}}
\def\gta{\mathrel{\spose{\lower 3pt\hbox{$\mathchar''218$}}
     \raise 2.0pt\hbox{$\mathchar''13E$}}}
\def\myputfigure#1#2#3#4#5%
{\hskip0.03\textwidth\vskip#5pt%\hfill
\makebox[0pt]{\hskip#2in
\includegraphics[width=#3\textwidth]{#1}}\vskip#4pt\hfill}
\newcommand{\beq}{\begin{equation}}
\newcommand{\eeq}{\end{equation}}


\lefthead{Haiman, Abel \& Madau}
\righthead{Minihalos and Reionization}

\begin{document}

\title{Photon Consumption In Minihalos during Cosmological Reionization}
\author{Zolt\'an Haiman\altaffilmark{1,2,6}, Tom Abel\altaffilmark{2,3} \& Piero Madau\altaffilmark{4,5}}
\affil{$^{1}$Princeton University Observatory, Ivy Lane, Princeton, NJ 08544, USA}
\affil{$^{2}$Institute for Theoretical Physics, University of California at Santa Barbara, CA 93106, USA}
\affil{$^{3}$Harvard-Smithsonian Center for Astrophysics, 60 Garden Street MS 10, Cambridge, MA 02138, USA}
\affil{$^{4}$Institute of Astronomy, Madingley Road, Cambridge, CB3 0HA, U.K.}
\affil{$^{5}$Department of Astronomy and Astrophysics, University of California, Santa Cruz, CA 95064, USA}

\altaffiltext{6}{Hubble Fellow.}

\authoremail{zoltan@astro.princeton.edu}
\authoremail{tabel@cfa.harvard.edu}
\authoremail{pmadau@ast.cam.ac.uk}

\vspace{\baselineskip}
\submitted{Accepted for publication in ApJ}

\begin{abstract}
At the earliest epochs of structure formation in cold dark matter (CDM)
cosmologies, the smallest nonlinear objects are the numerous small halos that
condense with virial temperatures below $\sim 10^4$K.  Such ``minihalos'' are
not yet resolved in large--scale three--dimensional cosmological simulations.
Here we employ a semi--analytic method, combined with three--dimensional
simulations of individual minihalos, to examine their importance during
cosmological reionization.  We show that, depending on when reionization takes
place, they potentially play an important role as sinks of ionizing radiation.
If reionization occurs at sufficiently high redshifts ($z_r \gsim 20$), the
intergalactic medium is heated to $\sim 10^4$K and most minihalos never form.
On the other hand, if $z_r\lsim 20$, a significant fraction ($\gsim10\%$) of
all baryons have already collapsed into minihalos, and are subsequently removed
from the halos by photoevaporation as the ionizing background flux builds up.
We show that this process can require a significant budget of ionizing photons;
exceeding the production by a straightforward extrapolation back in time of
known quasar and galaxy populations by a factor of up to $\sim 10$ and
$\sim 3$, respectively.

\end{abstract}

\centerline{{\it subject headings}: cosmology:theory -- early universe -- galaxies: evolution}

\section{Introduction}

The lack of any Gunn--Peterson troughs in the spectra of distant
quasars (Stern et al. 2000; Fan et al. 2000), as well as the presence
of Lyman $\alpha$ emission lines in the spectra of high--redshift
galaxies (Weymann et al. 1998; Hu et al. 1999), imply that the
hydrogen in the intergalactic medium (IGM) is highly ionized by
redshift $z\approx 6$.  It is widely believed that reionization was
caused by an early population of either galaxies or quasars. The
process of reionization and its impact on several key cosmological
issues have recently received much theoretical attention, and have
been studied by several authors using semi-analytic models (Meiksin \&
Madau 1993; Shapiro et al. 1994; Haiman \& Loeb 1997, 1998) and
three--dimensional numerical simulations (Gnedin \& Ostriker 1997).
In general, these works aim to follow the time evolution of the
filling factor of ionized (HII) regions, based on some input
prescriptions for the emissivity and spectra of the ionizing sources.

More recent works have focused on the increased rate of recombinations
in a clumpy medium relative to a homogeneous one, i.e. when $\langle
\rho^2 \rangle > \langle \rho \rangle^2$ (Ciardi et al. 2000; Benson
et al. 2000; Chiu \& Ostriker 2000; Gnedin 2000a; Madau, Haardt, \&
Rees 1999 [hereafter MHR]).  These studies have left significant
uncertainties on the details of how reionization proceeds in an
inhomogeneous medium.  Since the ionizing sources are likely embedded
in dense regions, one might expect that these dense regions are
ionized first, before the radiation escapes to ionize the low--density
IGM.  Alternatively, most of the radiation might escape from the
local, dense regions along low column density lines of sight.  In this
case, the underdense `voids' are ionized first, with the ionization of
the denser filaments and halos lagging behind (Miralda-Escud\'e et
al. 2000).

In this paper, we point out that that the density inhomogeneities at the
earliest redshifts are dominated by the smallest nonlinear structures,
i.e. halos near the cosmological Jeans mass, $M_{\rm Jeans}\approx 10^4~{\rm
M_\odot}$.  Such small scales have not yet been resolved in numerical
simulations of reionization (e.g. Gnedin 2000a), and have also not yet been
fully quantified in the semi--analytic works (although see Chiu \& Ostriker
2000 and Benson et al. 2000 for partial treatments).  Our main goals in this
work are (1) to quantify the importance of the high--redshift minihalos as
sinks of ionizing photons; and (2) to assess whether by extrapolating to early
times the known population of galaxies and quasars a sufficient number of UV
photons are produced for hydrogen reionization.

This paper is organized as follows.  In \S~2, we summarize the
ionizing photon budget from the known populations of galaxies and
quasars.  In \S~3, we describe our model of an individual
photoevaporating minihalo, based on three--dimensional numerical
simulations.  Combining this model with a hierarchical structure
formation scenario, in \S~4 we explicitly show that the ensemble of
minihalos dominate the clumping of the high--redshift IGM.  In \S~5,
we then compute the total number of ionizing photons consumed by the
photoevaporating minihalos in the same hierarchical cosmology.  In
\S~6, we argue that the minihalos must indeed have been
photoevaporated.  In \S~7, we discuss the relevance of the covering
factor of minihalos around ionizing sources, and the uncertainties
this implies for our results.  Finally, in \S~8, we offer our
conclusions and summarize the implications of this work.  Unless
mentioned otherwise, throughout this paper we adopt a flat
$\Lambda$CDM cosmology with $(\Omega_\Lambda,\Omega_{\rm
b},h,\sigma_8,n)=(0.7,0.04,0.7,0.9,1)$.

\section{Ionizing Photons from Known Sources}

The two most natural candidates for reionization are extensions of the known
populations of quasars, or of galaxies with ongoing star-formation, to
redshifts beyond $z>6$. Here we briefly review the number of ionizing photons
expected from these sources.  This will serve as a reference point for
comparison to the ionizing photon budget we obtain in our calculations below.

\subsection{Quasars}

The luminosity function (LF) of quasars has been measured in optical surveys,
and convenient parametric fitting formulae have been published by, e.g. Pei
(1995).  A more recent determination by the Sloan Digital Sky Survey extends
the measurement of the bright end of the LF to higher redshifts, and is
consistent with earlier optical results (Fan et al. 2000).  Here we adopt the
fitting formulae of MHR, which has a somewhat shallower slope towards higher
redshifts than, e.g., Pei (1995), and therefore predicts a larger number of
quasars when extrapolated to high $z$.  We have used this LF, together with the
average intrinsic spectrum of Elvis et al. (1995), to compute the production
rate of ionizing ($E>$ 13.6 eV) photons from quasars.  The faint-end slope of
the empirical LF is $d\log\Phi/d\log L=-1.64$, so that most ionizing photons
are produced by relatively bright quasars near the ``knee'' of the LF.  A
recent determination of the high--redshift LF in the soft X--ray band (Miyaji
et al. 2000) has yielded significantly weaker evolution out to $z\approx 4.5$
than seen in the optical.  Although the X--ray LF still has large uncertainties
at $z\gsim 2$, taking the X--ray results at face value may indicate the
existence of a larger abundance of high--redshift quasars than seen in the
optical.  If this inference is confirmed in future studies, it could increase
the number of ionizing photons produced by bright quasars relative to our
estimates below, unless this high--$z$ X--ray population is intrinsically
obscured at optical/UV wavelengths.

\subsection{Galaxies}

The total star formation rate in the universe can be estimated using the
sample of high--redshift galaxies found by the Lyman-break technique (Steidel
et al. 1999; Madau \& Pozzetti 2000).  Since the rest-frame UV continuum at
1500--2800 \AA\ (redshifted into the visible band for distant sources) is
dominated by the same short-lived, massive stars which are responsible for the
emission of photons shortward of the Lyman edge, the needed conversion factor,
about one ionizing photon for every 5 photons at 1500 \AA, is fairly
insensitive to the assumed IMF and is independent of the galaxy history for
$t\gg 10^{7.3}\,$ yr (MHR).  We normalize the number of ionizing photons to the
{\it observed} 1500 \AA\ flux (rest-frame), i.e. bypass the need for any
correction due to dust extinction (this is a good approximation provided the
mean color excess $E_{912-1500}=1.6E_{\rm B-V}$ is actually small).  We have
further assumed an average escape fraction of $f_{\rm esc}=50\%$ for the
ionizing radiation from the galaxy HI layers into the IGM, relative to the
escape fraction at 1500 $\AA$.  This value is a factor of $\sim 5$ higher than
that inferred in nearby starbursts (Leitherer et al. 1995).  Theoretically, the
escape fraction is expected to be {\it lower} at high redshifts, because of the
higher gas densities then (Wood \& Loeb 1999; Dove et al. 2000). However, a
recent measurement of the mean escape fraction from a sample of $29$ Lyman
break galaxies at $\langle z\rangle=3.4$ has yielded a value $\gsim 50\%$
(Steidel et al. 2000). Although the physical implications of this result are
still unclear (Haehnelt et al. 2000), to be conservative we adopted this value
here.

\myputfigure{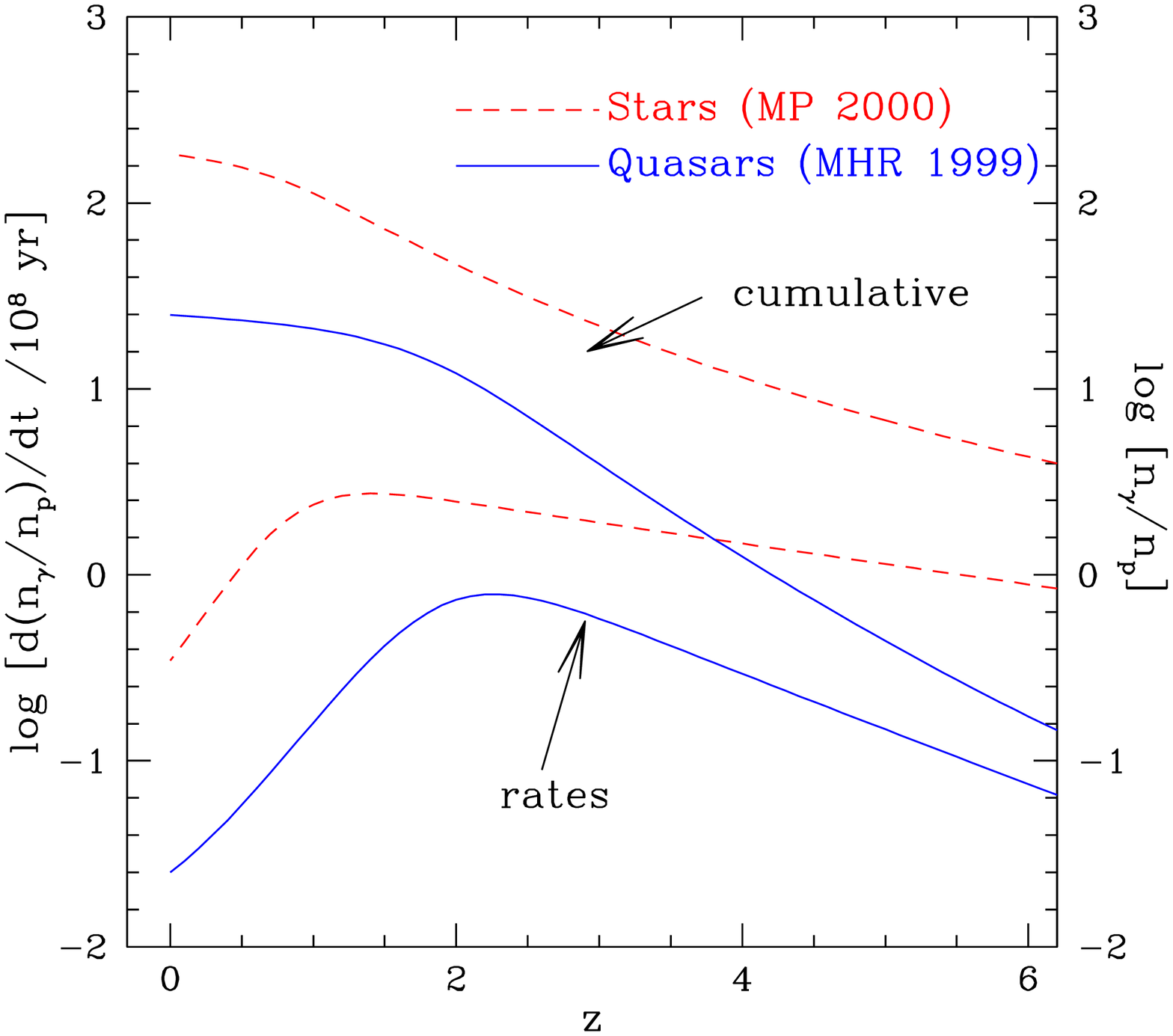}{3.2}{0.45}{-10}{-10} \figcaption{Production rate (lower
curves) and total number (integrated over cosmic time, starting from high
redshift, upper curves) of H-ionizing photons from known quasars and galaxies,
extrapolated to $z>4.5$.  The calculations are based on the counts of
high--redshift quasars and Lyman break galaxies.  We assumed the escape
fraction of ionizing photons to be $f_{\rm esc}=1$ for quasars and $f_{\rm
esc}=0.5\times f_{esc,1500}$ for galaxies. See text (\S~2) for discussion.
\label{fig:nion}}
\vspace{\baselineskip} 

The ionizing photon production rates for both quasars and galaxies are shown in
Figure~\ref{fig:nion}, in units of photons above 13.6 eV emitted into the IGM
per $10^8$ years per intergalactic hydrogen atom.  Also shown is the total
number of photons radiated per H-atom prior to redshift $z$.  To obtain
these integrals, we have extrapolated the photo emission rates $dn_\gamma/dz$
beyond $z=6$, using the observed slope $d\dot N/dz$ at $z=4$. These
extrapolations are clearly very uncertain.  For reference, we note that if the
photon emission rates had stayed flat at their $z=6$ values beyond $z>6$, than
in the stellar case, the total number of photons would increase very little
($\approx 20\%$), since the observed rate is already nearly constant in
redshift.  In the quasar case, this would result in an increase in the total
number of photons generated prior to $z=6$ by a factor of $\approx 4$.  Under
these extrapolations, as this figure reveals, luminous quasars produce little
ionizing radiation prior to redshift $z=5$, while galaxies provide $\approx
4\times (f_{\rm esc}/0.5)$ ionizing photons per hydrogen atom by this epoch.

\section{Photoevaporation of Minihalos}

In cold dark matter (CDM) cosmologies, the mean density and temperature of
the IGM imply the cosmological Jeans mass of

\begin{equation}
\label{eq:Mjeans}
M_{\rm J}\approx 10^4 
\left(\frac{\Omega_0h^2}{0.15}\right)^{-\frac{1}{2}}
\left(\frac{\Omega_{\rm b}h^2}{0.02}\right)^{-\frac{3}{5}}
\left(\frac{1+z}{11} \right)^{\frac{3}{2}}~{\rm M_\odot},
\end{equation}

where $M_{\rm J}$ is the total (gas + DM) mass of perturbations allowed to grow
in linear theory.  This mass corresponds to a virial temperature of the
dark matter halo

\begin{equation}
\label{eq:Tjeans}
T_{\rm J}\approx 25~
\left(\frac{h}{0.7}\right)^{2/3}
\left(\frac{\Omega_0h^2}{0.15}\right)^{-\frac{1}{3}} 
\left(\frac{\Omega_{\rm b}h^2}{0.02}\right)^{-\frac{2}{5}} 
\left(\frac{1+z}{11} \right)^2~{\rm K}.
\end{equation}

It is natural to expect that the earliest nonlinear objects are hosted in halos
of the size described by equations~(\ref{eq:Mjeans}) and (\ref{eq:Tjeans}).
Numerical simulations find that the mass scale can be up to a factor of $\sim
10$ still lower than this (Gnedin 2000b); however, as we shall find below our
results are insensitive to the exact value of the smallest allowed halo size,
as long as it is below $T\approx1000$K.  We shall hereafter refer to halos with
virial temperatures between $T_{\rm J} \leq T_{\rm vir} \leq 10^4$K as
``minihalos''.  A sufficiently strong early X--ray background (XRB) could
catalyze the formation of ${\rm H_2}$ molecules in minihalos, which could then
cool and fragment into stars, or form central black holes (Haiman, Rees \& Loeb
1996; Haiman, Abel \& Rees 2000).  In this case, the gas from the minihalos
would likely be ionized and expelled by the {\it internal} UV
source. Alternatively, in the absence of any early XRB, most minihalos would
likely not have sufficient molecular abundance to cool and dissipate (Haiman,
Rees \& Loeb 1997; Haiman, Abel \& Rees 2000).  In the further absence of any
{\it external} UV radiation, they would then remain in approximate hydrostatic
equilibrium until they eventually merge into more massive objects.

However, photoionization by either a nearby external UV source, or by
the smooth cosmic UV background radiation after the reionization
epoch, heats the gas inside a minihalo to a temperature $\approx
10^4$K.  By definition, the gas inside a minihalo is then no longer
bound, leading to the photoevaporation of the baryons out of their
host halos (Rees 1986; Shapiro, Raga \& Mellema 1998; Barkana \& Loeb
1999).  In general, the resulting photoevaporative outflow has a
complex velocity, density and ionization structure, whose
time--evolution has been studied in a two--dimensional simulation by
Shapiro \& Raga (2000).  In the present paper, we shall not address
this evaporation process itself in detail.  Our main goal here is to
obtain a rough estimate of how many recombinations each hydrogen atom
experiences during the evaporation process.  We will therefore use a
simplified model of the evaporative flow, which we calibrate using
three--dimensional simulations.

\subsection{Analytic Model of Photoevaporation}

We begin by adopting a radial density profile $n_{\rm H}(r)$ for the
undisturbed gas, based on the spherically symmetric truncated isothermal sphere
(TIS, Shapiro, Ilyev \& Raga 1999), with the scalings converted to our adopted
$\Lambda$CDM cosmology as given in Navarro, Frenk \& White (1997, hereafter
NFW).  This profile has the advantage of characterizing a spherical object with
a well--defined core, calibrated specifically for use within the
Press--Schechter formalism (which we shall use below).  For this density
profile, the average internal overdensity of the gas within the halo, relative
to the background hydrogen density, is $\delta_{\rm int}\equiv \langle n_{\rm
H} \rangle / \bar n_{\rm H} = 130.5$. Similarly, the mean ``internal clumping''
is $C_{\rm int}\equiv \langle n_{\rm H}^2 \rangle / \bar n_{\rm H}^2 = 444^2$.
As an alternative approach, we have assumed the gas to be in hydrostatic
equilibrium within a halo that has a density profile as parameterized by NFW,
with a concentration parameter $c=5$.  This value was obtained by following the
algorithm described in the appendix of NFW, which assigns to each halo of a
given mass identified at redshift $z$ a collapse redshift $z_{\rm coll}$
(defined as the time at which half of the mass of the halo was first contained
in progenitors more massive that some fixed fraction of the final mass), and
assuming that the characteristic halo density is proportional to the critical
density at $z_{\rm coll}$ (Navarro, Frenk \& White 1997).  With this algorithm
a minihalo of total mass $M=10^7\,h^{-1}\,$M$_\odot$ at $z=10$ has $(z_{\rm
coll},c)=(13.5, 4.8)$ (see also Madau, Ferrara, \& Rees 2000). The NFW profile,
by definition, has a mean internal overdensity of 200, and the embedded gas is
somewhat more centrally condensed than the TIS we adopted, $C_{\rm int}=532^2$.
We conclude that an NFW profile would give similar results to those obtained
below, except with the total number of recombinations increased by about
$(532/444)^2\approx 50\%$.

Within the core of the TIS profile, the central density contrast reaches a
value as high as $\sim 2\times 10^4$ relative to the cosmic background.  In
reality, for gas at temperature $T_{\rm vir}$ that has collapsed adiabatically,
the density contrast can not exceed $(T_{\rm vir}/T_{\rm IGM})^{3/2}$, where
$T_{\rm IGM}$ is the IGM temperature (given by $T_{\rm IGM}(z)\approx 2.73
(1+z_c) [(1+z)/(1+z_c)]^2$ with $z_c\approx 150$).  We therefore modify the
inner density profile, and adopt $\rho(r)={\rm min} [\rho_{\rm TIS}(r),
\rho_{\rm IGM}(T_{\rm vir}/T_{\rm IGM})^{3/2}] $, where $\rho_{\rm TIS}(r)$ is
the original TIS density run, and $\rho_{\rm IGM}$ is the density of the
uniform IGM.  This modification reduces the average density contrast and the
internal clumping below their fiducial values of 130.5 and 444 within halos
whose virial temperature is lower than $T_{\rm vir}\lsim 1500$K.  Below we
find that $\approx 68\%$ of the recombinations in the ensemble of minihalos
arise in those with $T_{\rm vir}\lsim 1500$K, so that uncertainties in the
density profiles of the smallest halos do not significantly effect our
results.
 
We next assume that the spherically symmetric halo is suddenly photoionized by
an external UV flux.  Simultaneously, the gas temperature jumps to $10^4$K,
resulting in an outflow.  We assume that photoevaporation requires a sound
crossing time, $t_{\rm pe}\approx R_{\rm vir}/10~{\rm km~s^{-1}}$, so that the
photoionized gas retains its original shape for a duration
\begin{equation}
\label{eq:tevap}
t_{\rm pe}= f \frac{R_{\rm vir}} {10~{\rm km~s^{-1}}}, 
\end{equation}
after which it is fully dispersed into the IGM (where $R_{\rm vir}$ is the
radius of the TIS).  Note that $t_{\rm pe}\propto T_{\rm vir}^{1/2}$, so that
smaller halos are photo--dissociated more rapidly. This prescription is
admittedly oversimplified, since the gas should expand gradually, and
non--uniformly, at varying speeds, rather than as a step--function.  In
addition, for low UV fluxes, the expansion might start before the gas is
ionized (see discussion below). Nevertheless, as we will see below, the
numerical simulations show a fairly rapid evaporation, and can be used to
calibrate the normalization constant ``$f$'' that we have introduced.

\subsection{Numerical Simulations}

Our main goal is to estimate the number $N_{\rm pe}$ of recombinations
each hydrogen atom experiences during the photoevaporation process.
This number is given by
\begin{eqnarray}
\label{eq:Nevap}
\nonumber N_{\rm pe} && = 
\int dt \int dV x^2 n_{\rm H}^2 \alpha_{\rm B} \times
\left[\int dV n_{\rm H} \right]^{-1}
\\
 && 
= f \frac{R_{\rm vir}} {10~{\rm km~s^{-1}}} C_{\rm int} \bar n_{\rm H} 
\alpha_{\rm B} / \delta_{\rm int}.
\end{eqnarray}
Here the volume integrals are taken over the region that contains the
virial gas mass, $x$ is the ionized fraction (which is a function of
radius and time), and $\alpha_B=2.6\times10^{-13}~{\rm cm^3~s^{-1}}$
is the case B hydrogen recombination coefficient evaluated at
$T=10^4$K.  Given an initial density profile, the second line can be
readily computed, except for the normalization constant $f$.  For
reference, we note that the photoevaporation time (eq.~\ref{eq:tevap})
for a $T_{\rm vir}=10^4$K minihalo at redshift $z=10$ is approximately
$25f\%$ of the Hubble time at that redshift; we find that during this
interval each hydrogen atom within this halo recombines $N_{\rm
pe}=340f$ times.

The first line in equation~(\ref{eq:Nevap}) is computed explicitly in the
simulation, which gives the total number of recombinations each hydrogen atom
experienced during the whole photoevaporation process.  By comparing this
number to the second line, we determine the normalization constant $f$.  The
simulations are performed by setting up stable truncated isothermal spheres in
both dark matter (DM) and gas as the initial condition for the
three--dimensional structured adaptive mesh refinement cosmological
hydrodynamics code {\it enzo} of Bryan \& Norman (1997; 1999).  For simplicity,
we use a standard CDM cosmology with $h=0.5$ and $\Omega_{\rm b}=0.1$.  We
solve the chemical reactions for the ions of hydrogen and helium, including
collisional ionization, radiative recombinations, and photo--ionization. The
reaction rates are taken from Abel et al. (1997).  The hydrogen
photo--ionization rate is assumed to be uniform on the grid(s); i.e. we do not
solve radiative transfer.  The photo--ionization heating rate is simply given
by $k_{\rm ph}\langle \epsilon_{\rm ph}\rangle$, where $\langle \epsilon_{\rm
ph}\rangle$ denotes the average energy of photo--electrons (which depends on
the ionizing spectrum; here we adopt $\langle \epsilon_{\rm ph}\rangle=2$eV).

The typical photoionization rates experienced by minihalos depend strongly
on the luminosity and clustering properties of the radiation sources.  In
addition, the photoionization rates can vary substantially for individual
minihalos, because of the $1/r^2$ dependence of the flux from nearby sources.
In our simulations, we have experimented with photoionization rates of $k_{\rm
ph}=10^{-13} - 10^{-10}~{\rm s^{-1}}$, corresponding roughly to a background
flux of $J=3\times10^{-23}-10^{-20}~{\rm erg~s^{-1}~cm^{-2}~Hz^{-1}~sr^{-1}}$
at 13.6eV.  In general, we find that for fast photoionization rates ($k_{\rm
ph}\gsim 10^{-12}~{\rm s^{-1}}$), the gas is fully ionized before any motion
occurs.  In this regime, we find a simple scaling of the numerical results (see
below) with halo mass and redshift, independent of the actual value of $k_{\rm
ph}$. This is because the heating--induced motions are independent of the
photo--heating rate, as long as this rate is comparable to or shorter than the
sound crossing time.  In the following, we will quote results from simulations
with a high--enough flux that allows us to use these scalings, and thus
eliminate the need for a large grid of simulations in mass, redshift, and flux.

We emphasize that for smaller UV fluxes, the number of recombinations can be
reduced relative to our adopted estimate. The core of the halo has a lower
ionization fraction, both because the equilibrium fraction is reduced, and
because the photoionization time scale in the core becomes comparable to the
sound crossing time (for $k_{\rm ph}\lsim 10^{-12}~{\rm s^{-1}}$), so that
equilibrium is not established before bulk motion occurs.  In addition,
radiative transfer effects for a low flux might keep the central regions
shielded until the density is significantly reduced, and further reduce the
number of recombinations.  In general, full radiative transfer would be needed
to adequately quantify these effects, which we defer to a subsequent paper.
Here we only estimate these effects in the optically thin case, and find that
for the range of $k_{\rm ph}$ of interest, the reduction in the number of
recombinations is small. As an example, for the low rate of $k_{\rm
ph}=10^{-13}~{\rm s^{-1}}$, the core of our fiducial $z=10$ halo would reach
$x=0.4$, resulting in an overall reduction of the internal clumping of the
ionized gas within the halo by a factor of two.  Finally, we note that prior to
reionization, it is conceivable that, due to the strong fluctuations of the UV
flux as a function of position, the gas condenses and photo--evaporates from DM
minihalos multiple times, and hence absorb many more photons.

Our calibrations are based on four simulations, using halo masses of
$M=10^6~{\rm M_\odot}$ or $M=5\times 10^6~{\rm M_\odot}$; and initial redshifts
of $z=10$ or $z=20$.  In all of these cases, by our choice of the flux,
photoevaporation occurs very rapidly, with most recombinations occurring within
a sound--crossing time; justifying the assumption of our simple model in
equation~(\ref{eq:tevap}). As an example, in Figure \ref{fig:simul} we show the
evolution of the $M=10^6~{\rm M_\odot}$ halo starting at $z=10$.  The figure
reveals that by $z=9.5$ the density contrast is everywhere reduced below
$\delta\lsim 10$. In each of the cases we examined, we infer $f=1$ from
equation~(\ref{eq:Nevap}) to within $\sim 15\%$. For the $M=10^6~{\rm M_\odot}$
halo, we find $f= 1.03, 0.86$ for $z=20$ and $z=10$, respectively.  Similarly,
for $M=5\times 10^6~{\rm M_\odot}$, we find $f= 0.85, 0.89$ for $z=20$ and
$z=10$.  Generally, considerable differences in the inferred values of $f$ can
be expected if one were to carry out radiative transfer calculations for more
realistic density distributions and source evolutions.  In the rest of this
paper, we take our four simulations at face value, and set $f=1$ in the
remaining calculations.

\myputfigure{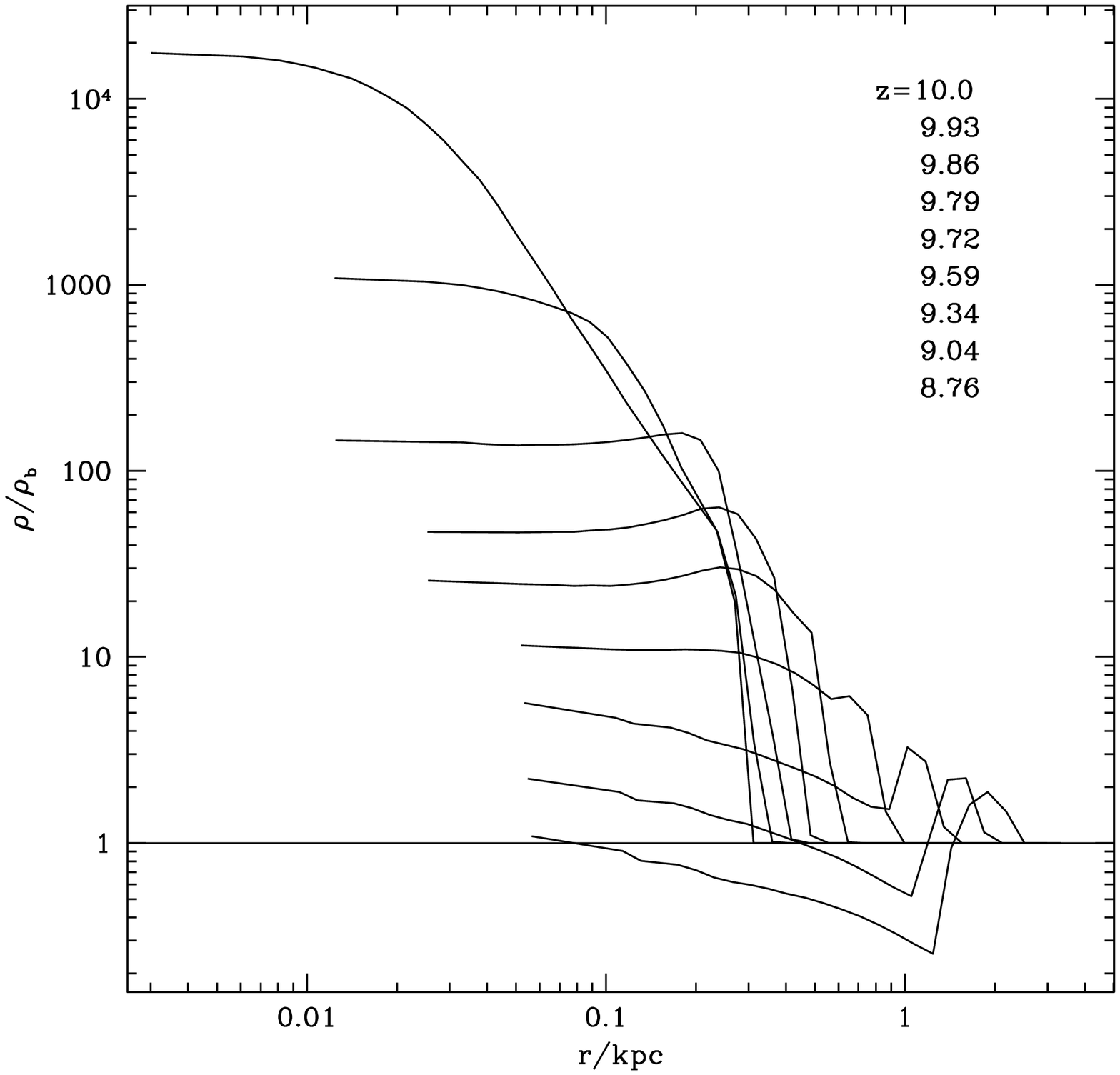}{3.2}{0.45}{-10}{-10} \figcaption{The evolution of the
density profile of a halo with mass $10^6~{\rm M_\odot}$, illuminated
by a background flux at $z=10$.
\label{fig:simul}}
\vspace{\baselineskip}

\section{Minihalos and Gas Clumping}

In this section, we illustrate the importance of minihalos by
estimating their contribution to the overall clumpiness of gas in the
universe.  For simplicity, we assume sudden reionization at some
redshift $z_r$, corresponding to the overlap of discrete HII regions,
and a sudden increase in the background flux at photon energy
$E=13.6$eV from zero to a value $J=J_{21}\times 10^{-21}~{\rm
erg~s^{-1}~cm^{-2}~Hz^{-1}~sr^{-1}}$. Prior to $z_r$ the background
flux is assumed to be negligible, so that minihalos form down to the
Jeans mass, or down to the virial temperature of $T_{\rm J}$
(eq.~\ref{eq:Tjeans}).

At the overlap epoch, the minihalos with $T_{\rm vir}<10^4$K contain a mass
fraction
\begin{equation}
\label{eq:Fcoll}
f_M(z_r) = F_{\rm coll}(z_r,T_{\rm J})-F_{\rm coll}(z_r,10^4~{\rm K})
\end{equation}
of all baryons, where $F_{\rm coll}(z,T)$ is the Press--Schechter
collapsed fraction at redshift $z$ above the mass cutoff corresponding
to the virial temperature $T$.  The minihalos therefore fill a
fraction
\begin{equation}
\label{eq:vfill}
f_V(z_r) \approx \frac{f_M(z_r)}{130.5}
\end{equation}
of the volume, where $\delta_{\rm int}=130.5$ is the average
overdensity within a minihalo relative to the background.  The
fraction we use in our calculations is somewhat larger than this value
because, as discussed in \S~3.1. above, we reduce the central
densities of the smallest minihalos relative to the TIS model; as a
result, they occupy a larger volume.

Once the ionizing flux turns on, the minihalos are photoionized, and
each minihalo contributes to the gas clumping and increases the
universal mean recombination rate, until it is photoevaporated.  To
illustrate the importance of minihalos, it is useful to define an
average gas clumping factor, 
\begin{equation}
C=\frac{\langle n_{\rm H}^2 \rangle}{\bar n_{\rm H}^2} = C_{\rm IGM}+C_{>}+C_{<}, 
\label{eq:cfac}
\end{equation}
where the three terms represent the contributions from the smooth IGM, halos
with $T_{\rm vir}>10^4$K, and minihalos, respectively. Previous works have
computed gas clumping in a similar manner, but have not included the
contribution $C_{<}$ from minihalos (see Benson et al. 2000).  Each term in
equation~(\ref{eq:cfac}) can be written as $C_i=f_{V,i}\langle n_i^2
\rangle/\bar n_{\rm H}^2$, where $f_{V,i}$ is the volume filling factor and
$n_i$ is the number density of ionized hydrogen in the $i^{\rm th}$ component.
The first term is obtained by assuming that the uncollapsed gas mass fraction
$\approx [1-F_{\rm coll}(z,T_{\rm J})]$ is fully ionized, and uniformly fills
the available volume $\approx [1-F_{\rm coll}(z,T_{\rm J})/130.5]$ (in our
calculations, we have included the small correction to these expressions due to
the reduction of the central densities of small minihalos).  Note that the
uncollapsed gas fraction is underdense relative to the overall mean of the
universe, so that $C_{\rm IGM}\leq 1$.

\myputfigure{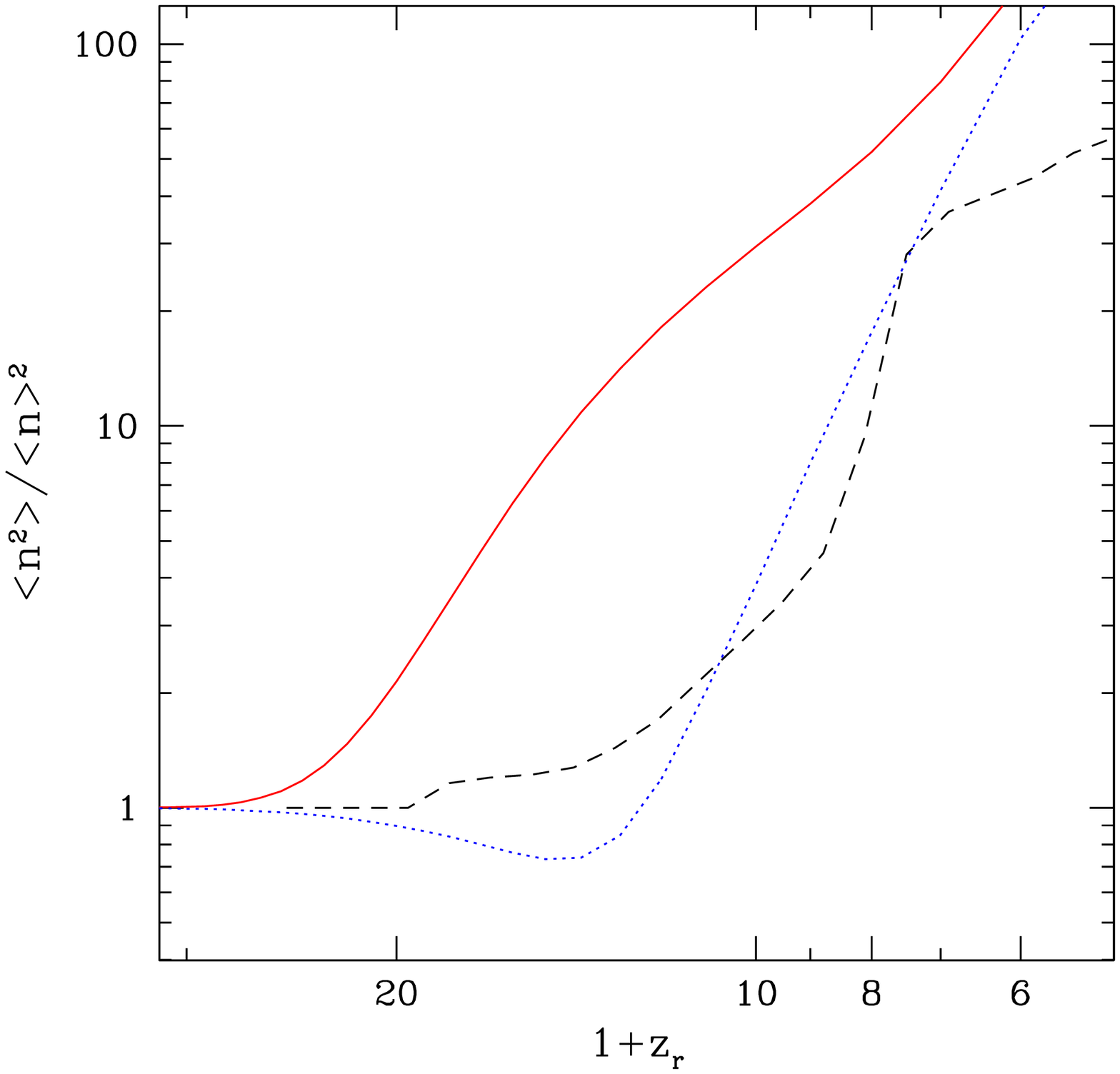}{3.2}{0.45}{-10}{-10} \figcaption{Average clumping
factor of ionized gas as a function of reionization redshift, when
minihalos are included (solid curve), and when only halos with $T_{\rm
vir}>10^4$K are included (dotted curve).  The dashed curve is from a
hydrodynamical simulation by Gnedin \& Ostriker (1997).
\label{fig:clump}}
\vspace{\baselineskip}

The contribution from large halos, $C_>$ is computed by summing over
all halos with $T_{\rm vir}>10^4$K, whose abundance is assumed to
follow the Press--Schechter mass function.  These halos are
collisionally ionized when they virialize, and subsequently cool via
atomic line cooling.  In the absence of an external UV field, most of
the hydrogen would recombine, and remain neutral.  When exposed to the
photoionizing background, the large halos are self--shielding, and are
still only partially ionized from the surface inward. This renders
estimates of the contribution from large halos to the overall clumping
highly uncertain, since the gas in each halo can continue to cool,
self-shield, and collapse to form a disk, while it is being
illuminated by the external radiation field.  Nevertheless, in order
to derive an illustrative estimate, we have computed the radius $r_i$
to which the halos are ionized from the outside in, assuming that the
gas density profiles follow that of the TIS solution.  We have assumed
further that the spherical halo is exposed to an external ionizing
flux of $J_{21}=1$, and repeated the 1D radiative transfer calculation
for several halo masses, to obtain $r_i$ as a function of $M_{\rm
halo}$.  We then evaluated $\langle n_i^2 \rangle$ for each halo mass,
excluding the neutral core $0<r<r_i$ in each case, and obtained the
final clumping factor by averaging $\langle n_i^2
\rangle$ over $M_{\rm halo}$.

Finally, the contribution $C_<$ from minihalos is computed by summing the
internal clumping $C_{\rm int}\leq 444^2$ over all minihalos with $T_{\rm
J}\leq T_{\rm vir}\leq 10^4$K, assuming that each minihalo is fully ionized,
but only exists for a time $t_{\rm pe}$ given by equation~(\ref{eq:tevap}).  To
illustrate the importance of minihalos, the resulting clumping factors $C_{\rm
IGM}+C_{>}+C_{<}$ and $C_{\rm IGM}+C_{>}$ are shown by the solid and dotted
curves in Figure~\ref{fig:clump}, respectively.  Also shown for comparison is
the clumping factor in a similar $\Lambda$CDM cosmology from the numerical
simulation of Gnedin \& Ostriker (1997, dashed curve).  Our calculations show
$C$ as a function of the assumed reionization redshift, while the simulation
result is adopted from a specific reionization history; this simple comparison
is therefore somewhat inappropriate.  Nevertheless, as the figure shows, the 3D
simulation results are well reproduced by the ensemble of partially ionized
spherical halos with $T_{\rm vir}>10^4$K, assuming a fixed background flux.
However, when the minihalos are included, clumping sets in at a significantly
higher redshift ($z\approx 20$ instead of $z\approx 10$).  Although the
relative importance of minihalos decreases at lower redshifts, when larger
scales start to collapse, the clumping factor is still enhanced by a factor of
$\sim 2$ at $z\approx 6$.

\section{Minihalos as Photon Sinks}

We will now explicitly compute the total number of ionizing photons
consumed inside minihalos following the overlap epoch $z_r$.  This
equals the number of recombinations that occur while the gas is
evaporated out of the minihalos.  Although larger halos with $T_{\rm
vir}>10^4$K eventually dominate the gas clumping, as emphasized in the
previous section, large halos can self--shield, cool, and collapse to
form disks, making it difficult to estimate their overall
contribution.  In order to avoid these complications, below we focus
on minihalos, and simply ignore halos with $T_{\rm vir}>10^4$K.

Consider a post--overlap redshift $z<z_r$, with $\Delta t(z,z_r)$
denoting the time elapsed between $z_r$ and $z$.  By this redshift,
the smallest minihalos have been evaporated, but those with virial
temperatures above

\begin{equation}
\label{eq:tempevap}
T_{\rm pe}(z)=10^4 \left[\frac{\Delta t(z,z_r)}{t_{\rm pe}(z_r,
10^4{\rm K})}\right]^2~{\rm K}
\end{equation}
are still present (cf. eq.~\ref{eq:tevap}) and are contributing to the
gas clumping.  The contribution of minihalos to the clumping factor at
redshift $z$ is accordingly given by

\begin{equation}
\label{eq:clump}
C_<(z)= \int_{M(T_{\rm min})}^{M(10^4{\rm K})} dM 
\left(\frac{dN}{dM}\right)_{z_r} 
\left( \frac{M}{\delta_{\rm int} \rho_{\rm bg}} \right)
C_{\rm int}
\end{equation}
where the first term in the integrand is the Press--Schechter mass
function; the second term represents the volume--filling factor of
halos of mass $M$; and the last term is the internal gas clumping of a
halo of mass $M$, $C_{\rm int}(M)\leq 444$.  The lower limit of the
integration corresponds to the smallest halo that has not yet been
photoevaporated, whose virial temperature is

\begin{equation}
\label{eq:Tmin}
T_{\rm min}(z) = {\rm max}(T_{\rm J},T_{\rm pe}).
\end{equation}

Since $T_{\rm J}$ is always evaluated at redshift $z_r$, the only
$z$--dependence of equation~(\ref{eq:clump}) are through the minimum
temperature of minihalos still being evaporated, and through the decrease of
the background density $\rho_{\rm bg}$.  Some of the minihalos are destroyed by
mergers; however, we find from the extended Press--Schechter formalism (Lacey
\& Cole 1993; eq. 28) that in the relevant ranges of redshifts (corresponding
to the photoevaporation time) and halo masses this fraction is always
small. The total number of recombinations per hydrogen atom in the universe
that takes places in minihalos between $z_r$ and $z$ is then given by

\begin{equation}
\label{eq:nrec}
N_{\rm rec}=\int_{t(z_r)}^{t(z)} dt \alpha_{\rm B} C_<(z) \bar n_{\rm H,0}
(1+z)^3,
\end{equation}
where $\bar n_{\rm H,0}$ is the present--day average density of hydrogen
atoms. Figure~\ref{fig:nrec} shows the evolution of $N_{\rm rec}$, assuming
different values for the reionization redshift, $6\leq z_r\leq 16$.  In
general, the number of recombinations associated with the photoevaporation of
minihalos is significant. For popular values for the reionization redshift, the
minihalos consume over $10$ ionizing photons per hydrogen atom.  Most
recombinations take place within a short time after reionization -- once the
minihalos are evaporated, they no longer contribute to recombinations, as
demonstrated by the rapid flattening of the curves to their asymptotic values
in Figure~\ref{fig:nrec}.  We recall that smaller halos are photoevaporated
more rapidly than larger ones, which makes their importance diminish towards
lower masses, despite their increasing abundance. We find that $\approx 68\%$
of the recombinations shown in Figure~\ref{fig:nrec} take place in those with
$T_{\rm vir}\lsim 1500$K; and excluding halos with $T_{\rm vir}< 1000$K would
make a negligible difference.  Our results imply that if the minihalos were
indeed photoevaporated (see discussion below), the ionizing sources must have
produced at least $\sim 10$ ionizing photons per background hydrogen atom. This
is a necessary, but not a sufficient condition for reionization (e.g. a sudden
burst of 10 ionizing photons per H atom could evaporate the minihalos and
reionize the universe, but would not ensure that most of the volume is
subsequently {\it kept} ionized).

\myputfigure{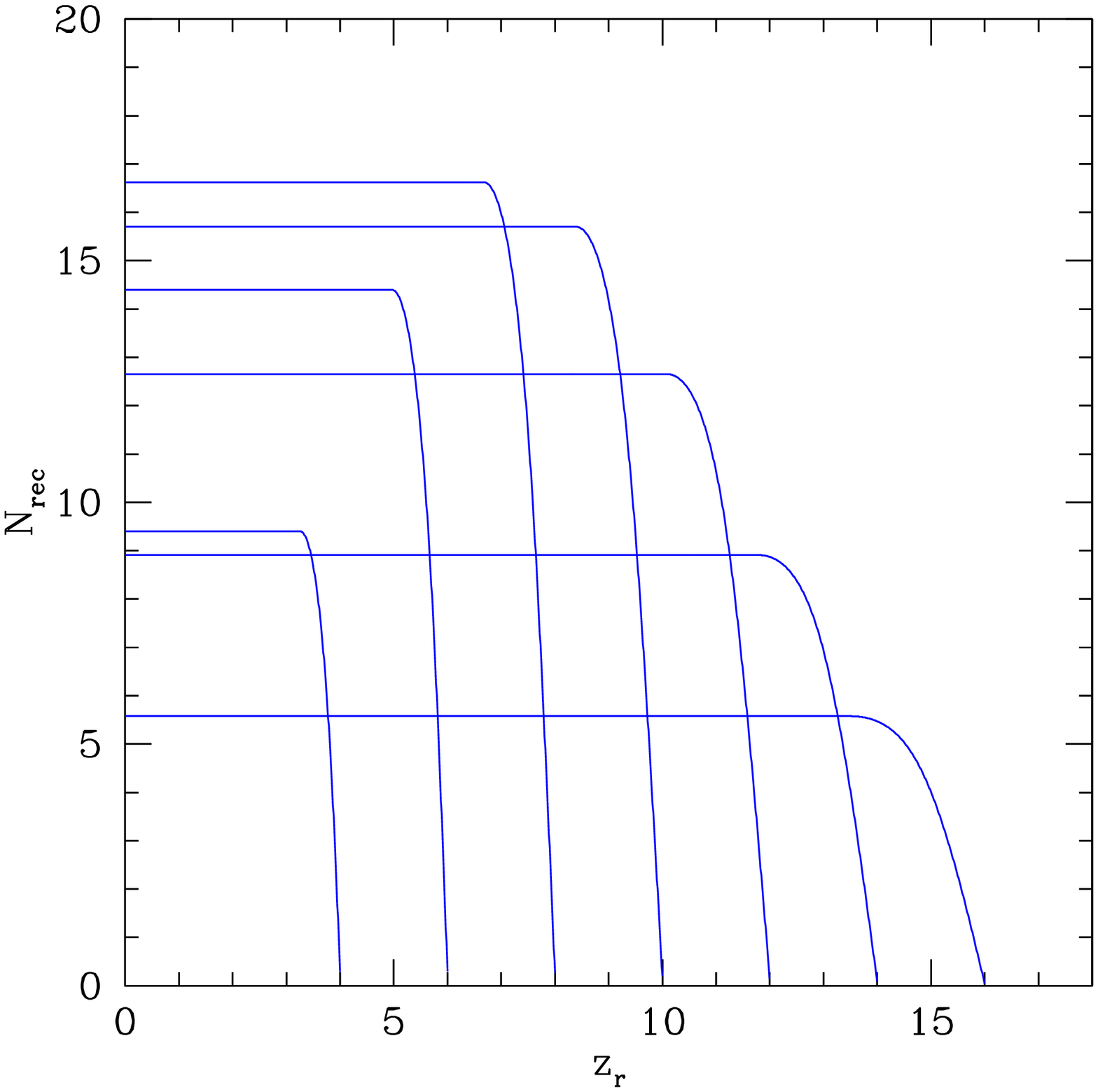}{3.2}{0.45}{-10}{-10} 
\figcaption{Total number of recombinations inside minihalos, following sudden
reionization at redshift $z_r$.  Recombinations are shown per hydrogen
atom, assuming the mean IGM density (see eq.~\ref{eq:nrec}).
\label{fig:nrec}}
\vspace{\baselineskip}

An interesting feature in Figure~\ref{fig:nrec} is the dependence of
the number of recombinations on the assumed reionization redshift.  If
reionization occurs early ($z_r\gsim 20$), i.e. by efficient ionizing
photon production in rare, high--$\sigma$ objects, the abundance of
minihalos is then still low.  Since we do not allow minihalos to
condense after reionization, most minihalos never form. As a result,
minihalos are less important sinks of ionizing radiation.  Likewise,
if reionization occurs late (closer to $z=6$), then the minihalos have
started merging into larger halos with $T_{\rm vir}>10^4$K.  In this
case, a fraction of minihalos reside in deep potential wells and can
be self--shielded, reducing their role as photon sinks.  The effect of
minihalos are maximal if reionization occurs around the popular values
of $z\approx 8-10$, the redshift when the minihalo abundance also
peaks (note that this redshift depends sensitively on the assumed
cosmology, and on the normalization of the power spectrum $\sigma_8$).

In our treatment, we assumed sudden reionization. The later stages of
reionization, corresponding to the slow outside--in ionization of
dense regions on larger scales, can last a considerable fraction of
the Hubble time. However, the initial overlap of HII regions likely
proceeds rapidly (Haiman \& Loeb 1998; Miralda-Escud\'e et al. 1999;
MHR; Gnedin 2000a).  Nevertheless, one could follow the volume filling
factor of ionized regions, and allow the formation of new minihalos
only outside these regions (as is done in Haiman \& Loeb 1998).  Here
we simply note that if the overlap epoch lasts between redshifts $z_1$
and $z_2$, then reading $N_{\rm rec}$ from Fig~\ref{fig:nrec} at these
two redshifts would approximately bracket the number of photons
consumed.

\section{The Mean Free Path of Ionizing Photons}

The minihalos that formed prior to the reionization redshift $z_r$
contribute to the Lyman-continuum opacity of the universe and reduce
the mean free path of ionizing photons at later redshifts $z<z_r$.  No
new minihalos form after reionization, and furthermore, a fraction
$1-f_{\rm surv}$ of the existing minihalos disappear by merging, as
they become parts of larger halos.  Assuming the minihalos to be
neutral and randomly distributed in space, the comoving photon mean
free path $d$ at redshift $z$ is given by

\begin{eqnarray}
\label{eq:mfp0}
\nonumber\frac{1+z}{d} && = \int_{M_{\rm J}}^{M_0} 
dM~n(M,z) \sigma(M) f_{\rm surv}(M,z_r,M_0,z)
\\
 && = \int_{M_{\rm J}}^{M_0} dM
\left(\frac{dN}{dM}\right)_{z_r} (1+z)^3 (\pi R_{\rm vir}^2) 
f_{\rm surv},
\end{eqnarray}
where $\sigma(M)$ is the geometrical cross--section, $R_{\rm vir}$ is
the virial radius, $n(M,z)$ is the physical abundance at redshift $z$,
and $(dN/dM)_{z_r}$ is the comoving abundance at redshift $z_r$ of
minihalos of mass $M$; and $f_{\rm surv}(M,z_r,M_0,z)$ is the
probability that a halo of mass $M$ at redshift $z_r$ has not become
part of a halo of mass $>M_0$ by redshift $z$.  Here and in the upper
limit of the integrals in equation~(\ref{eq:mfp0}), $M_0$ is chosen to
correspond to a virial temperature of $10^4$K.  The survival fraction
is computed as

\begin{equation}
\label{eq:fsurv}
f_{\rm surv}(M,z_r,M_0,z)= 1-\int_{M_0}^{\infty} dM_1
p(M,z_r,M_1,z),
\end{equation}
where $p(M,z_r,M_1,z)$ is the probability that a halo of mass $M$ at
redshift $z_r$ is part of a halo of mass $M_1\pm dM_1/2$ at redshift
$z$, given by Lacey
\& Cole (1993, eq. 28).

We find that the mean free path is small compared to the Hubble distance
$ct_{\rm Hub}(z)$.  For example, taking $z_r=10$ as the reionization redshift,
we find $d=2.1$(comoving) Mpc at $z=10$.  Equation~(\ref{eq:mfp0}) also shows
that the contribution to the opacity from halos of mass $M$ scales
approximately as $\propto M^{-1/3}$, implying that once the low--density IGM is
ionized, but the halos are still neutral, the minihalos are the dominant source
of opacity.  {\it As a result, the typical fate of an ionizing photon emitted
around redshift $z\approx 10$ is to be absorbed by a minihalo within a small
fraction of the Hubble distance.}

Although most ionizing photons are absorbed by minihalos, it is
interesting to follow the evolution of the mean free path at $z<z_r$
under the assumption that the minihalos stay neutral.  This could be
the case, e.g. if only $\sim$ one ionizing photon is produced in the
universe per H atom after $z_r$.  We find that approximately half of
the minihalos present at $z=10$ survive merging into large halos until
$z=3$, and the surviving minihalos lead to mean free paths of $d=10,
16,$ and $32$ Mpc at $z=5, 4,$ and $3$, respectively.  This implies
the existence of $dN/dz\approx 54, 42,$ and $31$ minihalos per unit
redshift at these redshifts.  The minihalos considered here have
hydrogen column densities of $N_{\rm H}\gsim 10^{18}~{\rm cm^{-2}}$,
and their abundances should therefore be compared to that of
high--redshift Lyman limit systems (LLS).  Although the latter is not
well measured at $z\approx 5$, it is inferred from high--resolution
quasar spectra to be $dN/dz\approx 2\pm 0.5$ at $z=3$ and $
dN/dz\approx 3.5\pm 1$ at $z=4$ (e.g. Storrie-Lombardi et al. 1994),
at least a factor of $\sim 10$ lower than the number of surviving
minihalos.  We conclude that the observed relatively low number of LLS
gives a strong constraint on the number of minihalos that could have
survived photoevaporation.  Unless future observations reveal a
10--fold increase in the abundance of Lyman limit systems from $z=4$
to $z=5$, this provides strong evidence that most minihalos were
indeed photoevaporated by $z\approx 5$ (or at least their total
geometrical cross--sections were reduced by an order of magnitude).
Indeed, Abel \& Mo (1998) have shown that halos with $T_{\rm vir}\gsim
10^4$K already have a sufficient abundance to account for all LLS at
$z\lsim 4$; leaving no room for additional minihalos.

\section{The Covering Factor of Minihalos}

So far we have assumed that reionization occurs suddenly.  Stated more
precisely, in this picture reionization occurs in three stages. First,
the HII regions around individual ionizing sources rapidly overlap and
fill most of the volume, occupied by the low--density IGM. Second, the
minihalos are photoevaporated by the background radiation. Finally,
the larger--scale dense regions are gradually ionized from the outside
inwards.  This scenario is justified as long as the minihalos have a
small covering factor $F_{\rm cov}$ around an ionizing source.  Here
$F_{\rm cov}$ is defined as the fraction of the $4\pi$ solid angle
around a typical ionizing source occupied by minihalos out to a
distance $r_s$, where $r_s$ is the typical separation of the ionizing
sources.  On the other hand, if $F_{\rm cov}\gsim 1$, then the
minihalos {\it must be} photoevaporated before the low--density
regions can be ionized, since ionizing photons are necessarily
absorbed in minihalos before they are able to reach most of the
volume.  In this case, the first two stages of reionization would be
reversed, and a typical minihalo is evaporated by a neighboring UV
source, rather than by a uniform background.  The resulting
``lopsided'' photoevaporation is considerably more complicated to
model than the evaporation by a uniform background, as we assumed in
\S~3.  Although such modeling is beyond the scope of the present
paper, we shall estimate the covering factor below.

To estimate the covering factor, let us assume that the ionizing
sources reside in halos with virial temperatures $T_{\rm vir}\approx
10^4$K, or mass $M_0$.  Taking $z_r=10$, the average separation of
these sources at the reionization epoch is
$r_s=[M_0(dN/dM_0)]^{-1/3}=0.5$ (comoving) Mpc.  If the minihalos were
randomly distributed, $F_{\rm cov}$ could be computed analogously to
the mean free path (cf. eq.~\ref{eq:mfp0}), and would be given simply
by

\begin{equation}
\label{eq:fcov0}
F_{\rm cov,0}=\frac{r_s}{d}\approx \frac{0.5}{2.1} \approx 25\%.
\end{equation}

However, equation~(\ref{eq:fcov0}) ignores the fact that minihalos are
strongly clustered around the ionizing sources, and therefore cover a
larger fraction of the sky.  Taking clustering into account, the
abundance of minihalos of mass $M_1$ at a distance $r\pm dr$ away from
an ionizing source in a halo of mass $M_0$ is given approximately by

\begin{equation}
\label{eq:clustering}
N\approx {N_1} [1+b_0 b_1 \xi_m(r)],
\end{equation}
where $N_1= M_1 (dN/dM_1)$ is the average minihalo abundance,
$\xi_m(r)$ is the mass correlation function at redshift $z=10$, and
$b_1=b(z_r,M_1)$ and $b_0=b(z_r,M_0)$ are the bias of halos of mass
$M_1$ and $M_0$ relative to mass.  The smallest spatial scale we
consider is the virial radius of a halo at the Jeans mass, $\approx 1$
(comoving) kpc, which is in the mildly non-linear regime. For
simplicity, we use the linear correlation function and halo bias,
although nonlinear effects could increase the clustering somewhat
above the estimates below. The relevant halo masses are always above
$M^*$, i.e. even the smallest minihalo has a positive bias parameter
$b(z,M)$.

For halos of a given mass $M_1$, clustering then increases the
covering factor to

\begin{equation}
\label{eq:fcov}
F_{\rm cov}= F_{\rm cov,0} [1+\frac{b_0 b_1}{r_s} \int_{R_{\rm vir}}^{r_s} dr
\xi_m(r)].
\end{equation}
Here $R_{\rm vir}$ refers to the virial radius of the minihalo of mass
$M_1$. Averaging the term in square brackets over all masses between
$M_{\rm J}$ and $M_0$, we find that the overall increase at redshift
$z=10$ is a factor of $\approx 3$.  The mass-dependence is primarily
through the bias parameter $b_1$, since the linear mass correlation
function $\xi_m(r)$ is quite flat, and the radial integral is
dominated by large radii.

In summary, our estimates imply a typical covering factor of $F_{\rm
cov}\approx 75\%$.  This relatively large value demonstrates that most ionizing
photons emitted by a source are absorbed in minihalos before most of the volume
of the IGM can be ionized.  Our estimates are inherently uncertain, as they are
based on spherical halos and the linear correlation functions. As higher
resolution three--dimensional simulations become feasible in the future, it
will possible to directly compute the covering factor of minihalos, and
eventually to model their asymmetric photoevaporation (as done in two
dimensions by Shapiro \& Raga 2000).

\section{Conclusions}

We have shown that minihalos, i.e. halos with virial temperatures between the
cosmological Jeans temperature, $T_{\rm J}$, and $10^4$K, play an important
role during cosmological reionization.  In the earliest stages of reionization,
the minihalos dominate the average gas clumping in the universe, and can
therefore dominate the overall recombination rate.  Depending on the
reionization redshift, up to $\approx 20\%$ of all baryons condense into
minihalos before their formation is suppressed by the reheating feedback.  The
fate of the chemically pristine minihalos depends on the abundance of ${\rm
H_2}$ molecules (see Haiman, Abel \& Rees 2000). A sufficiently strong early
X--ray background (XRB) could catalyze the formation of ${\rm H_2}$ molecules,
and the minihalos could then cool and harbor internal ionizing sources, likely
photo--evaporating their gas content from the inside (note that X--rays would
have several additional important consequences for reionization, see Oh 2000).
In absence of any early XRB, however, most minihalos would likely not have
sufficient molecular abundance, and they would then remain in approximate
hydrostatic equilibrium until photoevaporated by external ionizing sources.

We find that the photoevaporation of existing minihalos in the range of
virial temperatures $1000~{\rm K} \lsim T_{\rm vir} \lsim 10^4~{\rm K}$
dominates the necessary ionizing photon budget, and requires $10-20$ ionizing
photons per background hydrogen atom.  This value exceeds by about an order of
magnitude the number of ionizing photons produced by an extrapolation of known
populations of quasars to higher redshifts.  An extrapolation of the Lyman
break galaxy population to high redshift comes closer to meeting the required
photon budget, provided that a large fraction of the ionizing photons produced
in these galaxies leak into the IGM.  If the mean escape fraction is $\sim
50\%$ (cf. Steidel et al. 2000), then the discrepancy is only a factor of $\sim
3$.  Our conclusions depend further on the type of sources responsible for
reionization.  Our results show that if the reionizing sources have typical
separations of halos with virial temperatures of $10^4$K (or larger), then most
minihalos are photoevaporated before most of the volume of the universe is
reionized, because their covering factor around the ionizing sources is of
order unity.  If the reionizing sources are more closely packed, in principle
they could ionize the low--density regions, i.e. most of the volume, before the
minihalos are photoevaporated.  In either case, we find that the minihalos must
have been photoevaporated by $z\approx 5$, in order not to overpredict the
number of Lyman limit systems by a factor of $\sim 10$.

In a previous study of inhomogeneous reionization, Miralda-Escud\'e et
al. (2000) have suggested that $\sim$ one ionizing photon per hydrogen atom is
sufficient to reionize most of the volume of the IGM by $z\approx 5$.  The gas
clumping in that study was adjusted to match the results of numerical
simulations, which did not resolve the smallest scales, i.e. scales on which
minihalos dominate the gas clumping.  Our results suggest that the necessary
photoevaporation of dense gas out of minihalos can raise the photon requirement
at $z\gsim 5$ by up to an order of magnitude.  In arriving at this conclusion,
we have relied on a combination of a semi--analytic approach and
three-dimensional simulations of individual halos, albeit without radiative
transfer, and we have also made a number of simplifying assumptions. More
realistic cosmological hydrodynamical simulations that include realistic
sources and three--dimensional radiative transfer will be available in the near
future, and will be able to test our simplified model.

The additional ionizing photons required for the photoevaporation of
minihalos could arise from an early population of low--luminosity
quasars (``miniquasars''), whose abundance declines significantly less
than that of optically bright QSOs.  Alternatively, ionizing photons
could have been produced in ``minigalaxies'' associated with low--mass
halos collapsing at high redshift.  Although both types of sources
could have escaped detection with present instruments, both are well
within the capabilities of direct imaging with future telescopes, such
as the {\it Next Generation Space Telescope}.

%\acknowledgements
\vspace{0.5\baselineskip}

We thank Houjun Mo for useful discussions, and Jordi Miralda-Escud\'e, Martin
Haehnelt, Martin Rees, Rennan Barkana, and the referee, David Weinberg, for
useful comments on the manuscript.  We also thank Greg Bryan and Mike Norman
for permission to use their adaptive mesh refinement code {\it enzo}. This
research was supported by the NSF under Grant No. PHY94-07194 at the ITP, and
by NASA through the Hubble Fellowship grant HF-01119.01-99A, awarded to ZH by
the Space Telescope Science Institute, which is operated by the Association of
Universities for Research in Astronomy, Inc., for NASA under contract NAS
5-26555. PM acknowledges support by NASA through ATP grant NAG5--4236.

\end{document}